\documentclass[manuscript]{acmart}
\AtBeginDocument{%
  }

\usepackage{booktabs} 
\usepackage{enumitem}
\usepackage{amsmath}
\usepackage{amsthm}
\usepackage{amsfonts}
\usepackage{dsfont}
\usepackage{cleveref}
\usepackage{multirow}
\usepackage{marginnote}

\graphicspath{{media/}}     

\usepackage{xcolor}

\newcommand{\sw}[1]{\textcolor{black}{#1}}
\newcommand{\swmn}[2]{\textcolor{black}{{#2}}}

\newcommand{\x}{$\mathbb{X}$}
\newcommand{\xs}{$\mathbb{X}$\space}

\begin{document}

\title[Examining Community-Requested Fact-Checking]{Examining Community-Requested Fact-Checking: Request Alerts Are Associated with Greater Diversity and Visibility of Community Notes}

\author{Yilin Gong}
\email{yg24@iu.edu}
\orcid{0009-0003-6093-8922}
\affiliation{%
  \institution{Indiana University Bloomington}
  \city{Bloomington}
  \state{Indiana}
  \country{USA}
}

\author{Siqi Wu}
\email{swu2@iu.edu}
\orcid{0000-0002-1384-6908}
\affiliation{%
  \institution{Indiana University Bloomington}
  \city{Bloomington}
  \state{Indiana}
  \country{USA}
}

\begin{abstract}

Crowdsourced fact-checking systems such as Community Notes are increasingly used on social media platforms, yet concerns remain about which content receives scrutiny and how visible that scrutiny is. \xs allows users to request notes for specific posts. When sufficient requests accumulate, an alert is displayed, creating an interface cue that may guide contributor behavior. We present a quantitative, non-causal analysis comparing the diversity and visibility of community notes written for \xs posts with and without request alerts. We infer alert presence at note submission and analyze 10,432 alerted and 44,442 non-alerted English notes from 318 top writers. We find that, alerted notes are associated with greater individual-level topical diversity, but also with stronger collective concentration in the Politics category. Mixed-effects models estimate that alerted notes are 8.4-20.2 percentage points more likely to be modeled helpful and visible, though this visibility gain diminishes as topics diverge from writers' prior interests.

\end{abstract}




\received{20 February 2007}
\received[revised]{12 March 2009}
\received[accepted]{5 June 2009}

\maketitle

\section{Introduction}

Online misinformation threatens public discourse, distorts collective decision-making, and undermines trust in democratic institutions~\cite{del2016spreading,lazer2018science}. Confronting this pressing problem requires verification systems that are both timely and scalable, yet professional fact-checking remains constrained due to slow response times, low coverage rates, and high resource demands. In recent years, social media platforms have shifted toward crowdsourced fact-checking, such as Community Notes on \x~\cite{wojcik2022birdwatch} and similar designs on~\citet{youtube_cn}, \citet{meta_cn}, and~\citet{tiktok_footnotes}. Emerging evidence shows that Community Notes can effectively reduce the spread of and engagement with misleading content~\cite{slaughter2025community,gao2025can,chuai2024did,chuai2026community}, suggesting that crowdsourced approaches offer a promising complement to expert-led efforts to combat misinformation~\cite{allen2021scaling,martel2024crowds}.

Despite its promise, Community Notes faces two central criticisms. First, content is not equally checked. Contributors tend to write negative notes and provide negative ratings for content that contradicts their political views~\cite{allen2022birds}, with Republican posts receiving more scrutiny than Democratic ones~\cite{renault2025republicans}. Second, only a small fraction of submitted notes are ultimately displayed to the public~\cite{truong2025community}. Even notes that professional fact-checkers would deem helpful often fail to surface~\cite{ccdh}. Therefore, although prior research reveals meaningful corrective effects of Community Notes~\cite{slaughter2025community,kim2025differential,chuai2026community}, mitigating contributor biases in content selection and increasing the visibility of high quality notes remain critical priorities for future system improvement.

In the human-computer interaction (HCI) and communication literature, interface cues are widely recognized as effective interventions for shaping user perception, attention, and behavior~\cite{sundar2008main,metzger2010social,kim2011using}. Prior misinformation research also demonstrates that such cues can increase perceived accuracy and reduce sharing intention of false information~\cite{yaqub2020effects,lu2022effects,martel2024fact}. Recently, Community Notes introduced a feature that allows users to request fact-checking notes on specific \xs posts. When sufficient requests accumulate, \xs attaches an alert to the post,\footnote{A visual illustration of request alerts can be found here.~\url{https://communitynotes.x.com/guide/en/under-the-hood/note-requests}} signaling to potential note writers that this post requires additional scrutiny. This alert acts as a concrete interface cue, guiding writers' attention toward posts that the broader community has identified as needing context. \swmn{\shortstack{SPC-1a\\R3-1}}{In this work, we present a quantitative, non-causal study of request alerts based on public observational data, examining how the presence of alerts is associated with both the diversity of topics writers engage with and the visibility of the notes they produce.}

 Using the public Community Notes data, we first inferred the presence of request alerts at the time of note submission. We identified 318 top writers who repeatedly wrote notes for posts with alerts. We constructed two comparable groups of English notes: 10,432 written with request alerts and 44,442 written without, hereafter referred to as \textit{alerted} and \textit{non-alerted} notes, respectively. \swmn{\shortstack{SPC-1a\\R3-1}}{We use alert status (presence or absence) as the key comparison variable. To measure its association with post diversity,} we applied topic modeling to the concatenation of note texts for each \xs post and used axial coding to identify seven broad categories. We computed a writer's topical interests at any given time by using the embeddings of their recently checked posts and then derived a topic shift metric to quantify the movement of writer interests. To \swmn{\shortstack{SPC-1a\\R3-1}}{measure its association with} note visibility, we employed a mixed-effects model that accounted for both writer- and topic-level random effects to model the note publication status. We estimated the interaction effects of having an alert under different topical categories. We also examined the relationship between topic shift and note visibility.

This work contributes two new findings about request alerts, a community-driven interface cue design on an emerging fact-checking platform. First, notes written for posts with alerts more frequently deviate from writers' prior topical interests, specifically, shifting away from relatively soft topics (e.g., scams, science) and toward harder topics (e.g., politics, social issues). Second, alerted notes are more likely to be modeled as helpful and shown to the public. The differences are statistically significant and consistent across all categories, ranging from 8.4 percentage points in Politics \& Conflict to 20.2 percentage points in Science \& Technology. However, this visibility gain diminishes as topics diverge further from writer's prior interests. \swmn{\shortstack{SPC-1a\\R3-1}}{These findings demonstrate how contributor activity and note visibility vary in relation to community requests,} and highlight design considerations for supporting more equitable and effective crowdsourced misinformation governance.


\section{Background and Related Work}

\subsection{The Community Notes Ecosystem}
\label{ssec:background}

\x's Community Notes is one of the most prominent implementations of crowdsourced fact-checking on social media. Originally launched as Birdwatch~\cite{wojcik2022birdwatch}, the system allows users to write and rate contextual notes on potentially misleading \xs posts. Note helpfulness is determined through cross-partisan consensus, also known as bridging algorithms~\cite{ovadya2023bridging}, rather than simple majority voting. Notes can have one of three modeled publication statuses: currently rated helpful (CRH), currently rated not helpful (CRNH), or needs more ratings (NMR). Only notes modeled as CRH are visible to users on \x.

Community Notes has attracted substantial scholarly and public attention, in part due to its high level of transparency. \xs releases detailed documentation and the codebase of the underlying bridging algorithm, as well as up-to-date, real-world datasets containing note texts, ratings, and status histories. This openness has enabled independent audits and empirical evaluations of the system's performance, bias, and impact~\cite{allen2022birds,chuai2026consensus}. For example, prior work shows that when notes are displayed, they can reduce the engagement with misleading content, including decreasing resharing~\cite{slaughter2025community} and increasing the likelihood that authors delete posts~\cite{chuai2026community}. Motivated by its perceived legitimacy and scalability, other major platforms including \citet{youtube_cn}, \citet{meta_cn}, and \citet{tiktok_footnotes}, have recently begun experimenting with Community Notes-like systems for fact-checking. However, researchers have also identified systematic political biases in which content receives notes and ratings~\cite{allen2022birds}, as well as a low fraction of visible notes~\cite{ccdh,truong2025community}. These findings suggest that Community Notes can exert meaningful corrective effects, but its impact is limited by contributor selection biases and low note visibility.

\swmn{SPC-2}{On July 18, 2024, Community Notes introduced a request feature that allows users to explicitly indicate that a post may benefit from additional context.\footnote{\url{https://x.com/communitynotes/status/1813980126117609624}} Initially, when a post received at least five requests, \xs attached an alert to the post, signaling community interest. The criteria for displaying request alerts have since changed several times~\cite{github_requests}. On September 5, 2024, the threshold was revised to be at least $\max(5, \text{\#views}/25{,}000)$ requests, making alerts harder to trigger for highly viewed posts. On June 26, 2025, the fixed component of the threshold was lowered from five to four requests, relaxing the requirement for less-viewed posts. Most recently, on September 5, 2025, Community Notes introduced a \texttt{requestor helpfulness score}, allowing helpful users to trigger alerts with fewer requests.}

\sw{Request alerts are visible only to top writers and remain visible for 24 hours. Community Notes currently defines top writers as writers who meet two criteria: (i) at least 4\% of submitted notes are currently modeled as CRH and (ii) writing impact (defined as \#CRH minus \#CRNH) is at least ten~\cite{cn_top_writers}. In addition to seeing posts with alerts on the main \xs interface, top writers can access a dedicated ``Requests'' tab in their Community Notes dashboard. Besides, top writers are awarded a digital badge and their notes are more likely to trigger notifications to attract rater's attention.}

These features mark a shift from purely self-selected note creation toward a more demand-oriented process. Conceptually, request alerts serve as an interface cue for surfacing posts that users collectively perceive as ambiguous, misleading, or in need of clarification. While prior work has largely focused on the downstream effects of published notes, little attention has been paid to the upstream role of community requests. To our knowledge, \citet{chuai2026request} and \citet{pilarski2026supply} are the only studies that examine the request mechanism on Community Notes. \citet{chuai2026request} characterize notes created in response to requests, while \citet{pilarski2026supply} reveal the misalignment between community requests (demand) and notes (supply). In contrast, \swmn{\shortstack{SPC-1a\\R3-1}}{our work examines how request alerts are associated with the diversity and visibility of community notes.}

\subsection{Interface Cues}

Interface cues are design elements embedded in online platforms that signal information, direct attention, and shape user behavior. These cues often function as visual, structural, or contextual signals, and influence how users interpret and interact with digital content. Prior research conceptualizes interface cues as interventions that leverage heuristic processing~\cite{sundar2008main}, allowing users to rely on cognitive shortcuts when evaluating credibility, relevance, or social consensus in complex information environments~\cite{kim2011using,lin2016social}. As a result, interface cues play a crucial role in mediating user interactions with online systems, shaping both individual judgments and collective behavior at scale.

Previous research has identified several types of interface cues that shape user cognition and behavior through distinct mechanisms. Credibility cues, such as verification badges, source labels, and fact-check labels, influence users' judgments about the trustworthiness and accuracy of information~\cite{yaqub2020effects,martel2024fact}. Social cues, including popularity indicators and peer feedback signals, convey social consensus and can affect perceptions of relevance~\cite{kim2011using}. Attention cues, such as visual highlighting, alerts, and ranking positions, guide users' focus toward specific content by increasing salience. Scholars further conceptualize these cues within heuristic frameworks, most notably Sundar's MAIN model, which posits that modality, agency, interactivity, and navigability cues activate cognitive heuristics that users rely on when processing information in complex digital environments~\cite{sundar2008main}. This rich body of work demonstrates that interface cues systematically shape how users attend to, interpret, and act on online content, making them a crucial design mechanism in sociotechnical systems. 

Using this taxonomy, request alerts in Community Notes can be understood as a hybrid interface cue that combines attention and social signaling. As an attention cue, request alerts increase the salience of specific posts by signaling to potential contributors that a post has been flagged by the community for additional scrutiny~\cite{kim2011using}. At the same time, they function as a social cue by conveying aggregate signals of collective concern, implicitly indicating that multiple users perceive the post as ambiguous or potentially misleading~\cite{metzger2010social}. Unlike credibility cues, which directly affect perceived accuracy~\cite{pennycook2020implied,martel2024fact}, request alerts intervene earlier in the fact-checking process by influencing which content receives attention and who chooses to contribute. This positions request alerts as a distinctive interface-level mechanism for coordinating effort in crowdsourced fact-checking systems. 


\section{Community Notes and Requests Data}
\label{sec:data}

We downloaded three specific datasets from the public Community Notes data repository\footnote{\url{https://x.com/i/communitynotes/download-data}} on June 27, 2025, which include contributions up until June 25, 2025.

\begin{itemize}
    \item \textit{Notes dataset.} It contains metadata for 1,962,919 available community notes, including note ID, writer ID, \xs post ID, submission timestamp, and note text. \sw{Deleted notes, as well as notes associated with unavailable posts, are not included in this dataset, regardless of whether they were removed by users or by the platform.}
    \item \textit{Note Status History dataset.} It contains the publication status history of 2,108,340 notes, including every status change and its associated timestamp. \sw{This dataset retains notes that were previously modeled, even if they are no longer present in the Notes dataset.}
    \item \textit{Note Request dataset.} It contains metadata for 6,314,903 community requests, including requester ID, \xs post ID, and submission timestamp. 165,788 \xs posts have at least five requests, of which 95,807 posts have at least one note.
\end{itemize}

\paragraph{Inferring historical note, writer, and request alert status}

\begin{figure}[tbp]
    \centering
    \includegraphics[width=0.6\linewidth]{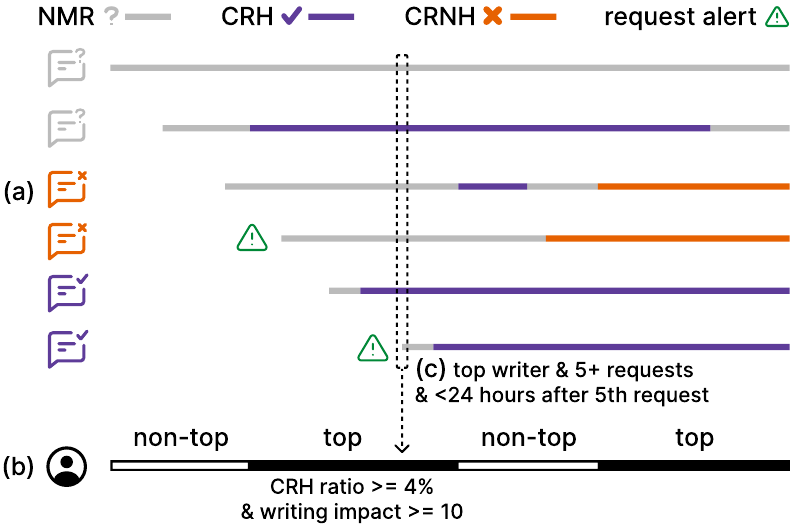}
    \caption{Illustration of (a) note publication status timelines, (b) top writer status timeline, (c) inferred request alert status at the time of note submission.}
    \Description{Illustration of (a) note publication status timelines, (b) top writer status timeline, (c) inferred request alert status at the time of note submission.}
    \label{fig:data_illustration}
\end{figure}

We used the Note Status History dataset to reconstruct the publication status timeline for each note. All notes are initialized with NMR status. The Community Notes bridging algorithm continuously updates note statuses as new ratings are received. In our dataset, 83.1\% of notes remain in the initial NMR status throughout their lifecycle; 11.3\% turn to either CRH or CRNH and stay at it; the remaining 5.6\% transition through more than two statuses. We consider two alternative metrics of note visibility: currently CRH and ever CRH. The former indicates whether a note's current publication status is CRH at the time of data collection, while the latter indicates whether the note has ever been modeled as CRH. \Cref{fig:data_illustration}(a) illustrates the publication status timelines for six notes. Among these, two notes are currently modeled as CRH (bottom two), while four notes have been modeled as CRH at some point (the second, third, and bottom two).

For each writer and at any given timestamp, we calculated the respective cumulative numbers of CRH, CRNH, and NMR notes they had contributed up to that point (see \Cref{fig:data_illustration} dashed vertical box). We then inferred their top writer status based on the criteria described in~\Cref{ssec:background}. \Cref{fig:data_illustration}(b) illustrates a writer who transitions multiple times between top writer and non–top writer status, as decided by the evolving publication statuses of their submitted notes over time. Overall, 15.4\% of notes were estimated to be written while their writers held top writer status.
    
We then inferred whether a request alert was displayed at the time of note submission. Notes written by top writers on \xs posts that received at least five requests and submitted within 24 hours of the fifth request were classified as having been created under a request alert, as illustrated in~\Cref{fig:data_illustration}(c). Using these three criteria, we identified 14,827 alerted notes from 1,034 top writers. 

\swmn{SPC-2}{Between September 5, 2024 and June 25, 2025 (the date of our data collection), highly viewed \xs posts were subject to a higher alert threshold: at least $\max(5, \text{\#views}/25{,}000)$ requests. Because historical view counts for \xs posts are not retrospectively available, we cannot fully verify whether each post satisfied the view-adjusted threshold at the time a note was submitted. We thus caution that our set of 14,827 identified alerted notes may include some false positives. To assess the extent of this concern, we calculated the number of requests associated with each identified alerted note at the time of submission. The mean value is 21.9 and the median is 8. This suggests that most identified alerted notes had request counts well above the minimum five-request threshold.} 

The absence of historical view counts has prompted prior studies to use different approaches for inferring request alert status. In contrast to our method, \citet{chuai2026request} rely solely on the number of requests and the final CRH ratio. The approach used by~\citet{pilarski2026supply} is more similar to ours, as they also take into account writers' historical writing impact scores, dynamic CRH ratios, and the duration for which alerts are visible. \citet{pilarski2026supply} calculate an estimated post view count to address the missing historical view count problem. We choose not to adopt this approach because it requires collecting additional \xs post data, which is prohibitively expensive, introduces further data attrition, and compromises reproducibility.

\paragraph{Data processing}
The 1,034 top writers had contributed a total of 311,467 non-empty notes. First, because we need to apply topic modeling to measure writers' topical interests in~\Cref{ssec:diversity_methods}, we restricted our analyses to English content for better modeling results. We used \texttt{Lingua}\footnote{\url{https://pypi.org/project/lingua-language-detector}} to detect the note language and included only the 191,011 notes classified as English. \sw{Second, because our data span nearly one year, we extracted all notes written between each writer's first and last alerted note to compare alerted and non-alerted notes within writers while minimizing potential temporal confounds.} We additionally included five non-alerted notes written before the first alerted note to provide pre-alert observations. 

\begin{figure}[t]
    \centering
    \includegraphics[width=1\linewidth]{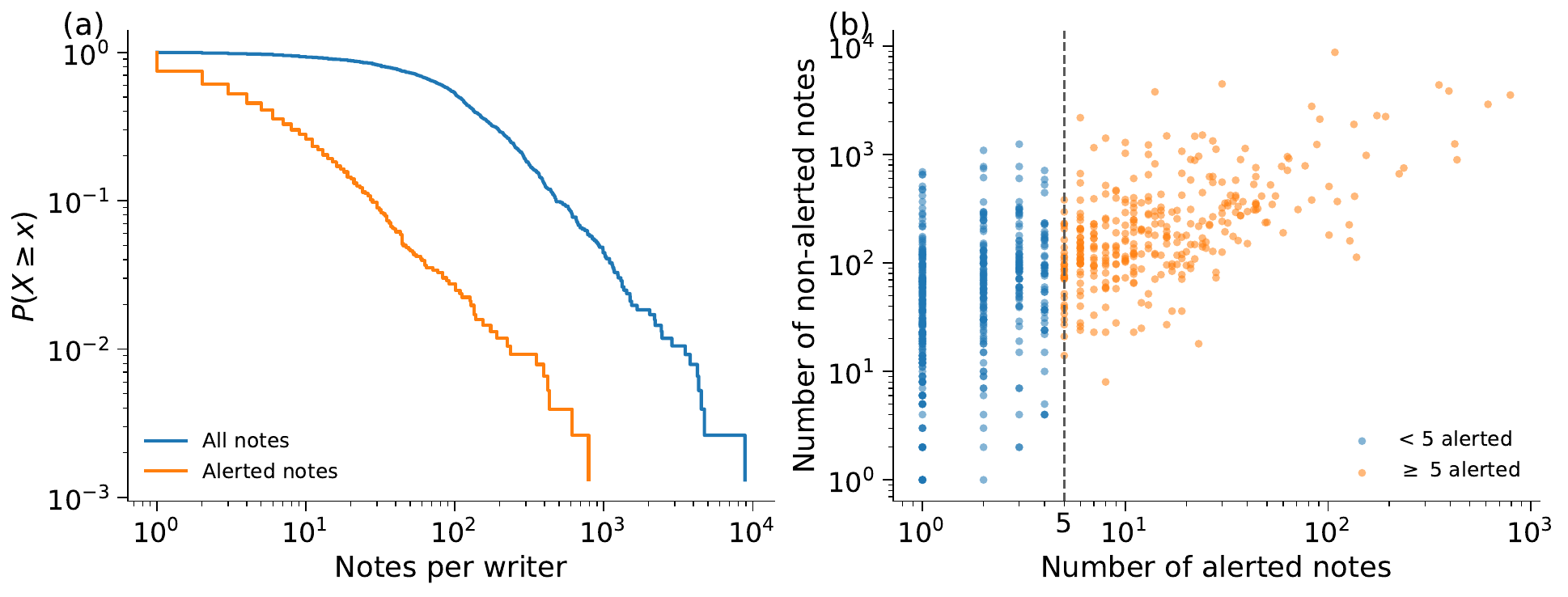}
    \caption{(a) CCDF of notes and alerted notes per writer. (b) Scatter plot of alerted vs. non-alerted notes per writer. The dashed vertical line marks the five-alerted-note threshold.}
    \Description{(a) CCDF of notes and alerted notes per writer. (b) Scatter plot of alerted vs. non-alerted notes per writer. The dashed vertical line marks the five-alerted-note threshold.}
    \label{fig:data_desc}
\end{figure}

\swmn{R3-3}{\Cref{fig:data_desc}(a) presents the complementary cumulative density function (CCDF) of notes and alerted notes per writer on a log-log scale. We observe that both distributions exhibit a highly skewed pattern, with many writers contributing relatively few notes and a small number of writers contributing substantially more. \Cref{fig:data_desc}(b) plots the number of alerted notes against the number of non-alerted notes for each top writer. The two quantities are positively correlated. To ensure reliable repeated measures at the writer level, we excluded writers with fewer than five alerted notes (i.e., those located to the left of the dashed vertical line in \Cref{fig:data_desc}(b)). Applying these criteria yielded 318 active top writers. Although these filters retains only 30.8\% of top writers exposed to request alerts, it preserves 70.4\% of alerted notes, providing substantial coverage of the alerted note sample.}


The final dataset consists of 318 top writers and two comparable groups of notes written on 51,089 \xs posts over similar time periods. In total, these writers contributed 54,874 English notes. The alerted group has 10,432 (mean: 32.8 per writer, median: 13) notes, while the non-alerted group has 44,442 (mean: 139.8, median: 52) notes. 
For brevity, we hereafter refer to top writers simply as writers, as our analyses focus exclusively on this population.


\section{Association with Topical Diversity}
\label{sec:diversity}

Here we first describe a topic modeling pipeline and an axial coding procedure for learning high-dimensional embeddings, fine-grained topics, and broad categories for \xs posts. We define a new metric, called ``topic shift'', to quantify how the topics of posts for which a writer submits notes diverge from their prior contributions. Next, we analyze the diversity of notes written with and without request alerts at both the individual and collective levels.

\subsection{Methods}
\label{ssec:diversity_methods}

\paragraph{Topic modeling}
The topics of the \xs posts that writers choose to fact-check reflect their interests. Accordingly, our goal is to learn the topic embeddings of \xs posts for which writers have submitted notes. Our processed dataset contains 54,874 notes for 51,089 \xs posts. To obtain richer texts for topic modeling, we referred to the original Community Notes dataset and extracted all English notes associated with these posts, without restricting writers to the 318 active top writers. For each post, we concatenated the texts of its associated notes into a single document. The resulting corpus contains 101,298 English notes, averaging 2.0 notes per post. 22,022 (43.1\%) posts have more than one note.

\swmn{SPC-3}{We use concatenated-note embeddings to represent the topics of \xs posts. While embedding the original post would provide the most direct representation, post content is not included in the public Community Notes dataset. Collecting \xs data has become prohibitively expensive, and many posts contain images or videos that are not well captured by text-only embeddings. An alternative is to embed only a single note, but it is also undesirable because a single note captures writer-specific framing and writing style in addition to the post's topic. Concatenating all notes associated with a post provides a richer and more robust proxy for post content by incorporating multiple perspectives on the same post. This choice is also consistent with the topic modeling module implemented in the official Community Notes codebase~\cite{github_topic_modeling}. As a robustness check, reported in~\Cref{app:post_embedding}, we show that cosine distances calculated from concatenated-note embeddings are significantly positively correlated with those calculated directly from post embeddings on a sample of \xs posts (Spearman's $\rho = 0.567, p<0.001$, Pearson's $r = 0.534, p<0.001$), supporting the use of concatenated-note embeddings as a proxy for post content.}

Our topic modeling pipeline is inspired by~\citet{kim2025differential}. We first removed URLs from the corpus. We then used \texttt{Twitter4SSE}~\cite{di2021exploiting} to encode the concatenated-note texts into 768-dimensional embeddings. Next, we applied UMAP for dimensionality reduction and HDBSCAN for clustering~\cite{mcinnes2017hdbscan}. This yielded 239 initial topic clusters. For each cluster, we applied class-based TF-IDF to extract the top ten keywords and three representative notes. 

\begin{table*}[t]
\centering
\small
\begin{tabular}{lr|rrr|rr}
\toprule
Broad Category & Representative Keywords & \multicolumn{3}{c}{Note Count} & \multicolumn{2}{c}{Ever CRH Rate} \\
 &  & alerted & non-alerted & total & alerted & non-alerted \\
\midrule
C1: Politics \& Conflict & Trump, voting, war, Gaza & 4,862 & 17,017 & 21,879 & 23.26\% & 13.50\% \\
C2: Policy Violation & gambling, altered, clickbait, fake & 2,260 & 11,113 & 13,373 & 36.15\% & 17.62\% \\
C3: Social \& Civil Issues & protest, crime, gang, immigrant & 1,402 & 5,303 & 6,705 & 26.89\% & 15.27\% \\
C4: Health & COVID, autism, vaccine, disease & 426 & 4,152 & 4,578 & 23.47\% & 20.45\% \\
C5: Science \& Technology & NASA, energy, food, chromosome & 552 & 3,387 & 3,939 & 42.39\% & 20.64\% \\
C6: Entertainment & artist, NBA, game, Olympics & 582 & 2,105 & 2,687 & 35.05\% & 21.28\% \\
C7: Business \& Finance & stock, market, tariff, bitcoin & 348 & 1,365 & 1,713 & 24.14\% & 12.75\% \\
\midrule
Total & & 10,432 & 44,442 & 54,874 & 28.25\% & 16.28\% \\
\bottomrule
\end{tabular}
\caption{Breakdown of note counts and ever CRH rates by broad category and request alert status, sorted by note counts. Ever CRH rates show the proportion of notes that hold CRH status at some point, even if they later lose it.}
\label{tab:broad_category}
\end{table*}

These clusters capture fine-grained topics such as election fraud, Russia-Ukraine war, and Gaza-Israel conflict. While informative, they do not constitute an exhaustive set. For instance, future posts may address previously unseen issues as they emerge over time, necessitating the creation of new topics. Nevertheless, they all belong to a broad Politics category. Because the mixed-effects models in~\Cref{ssec:visibility_methods} require a comprehensive and stable set of note content labels to include as fixed effects, we apply an axial coding procedure to group fine-grained topics into a fixed set of broad categories.

\begin{figure}[t]
    \centering
    \includegraphics[width=0.6\linewidth]{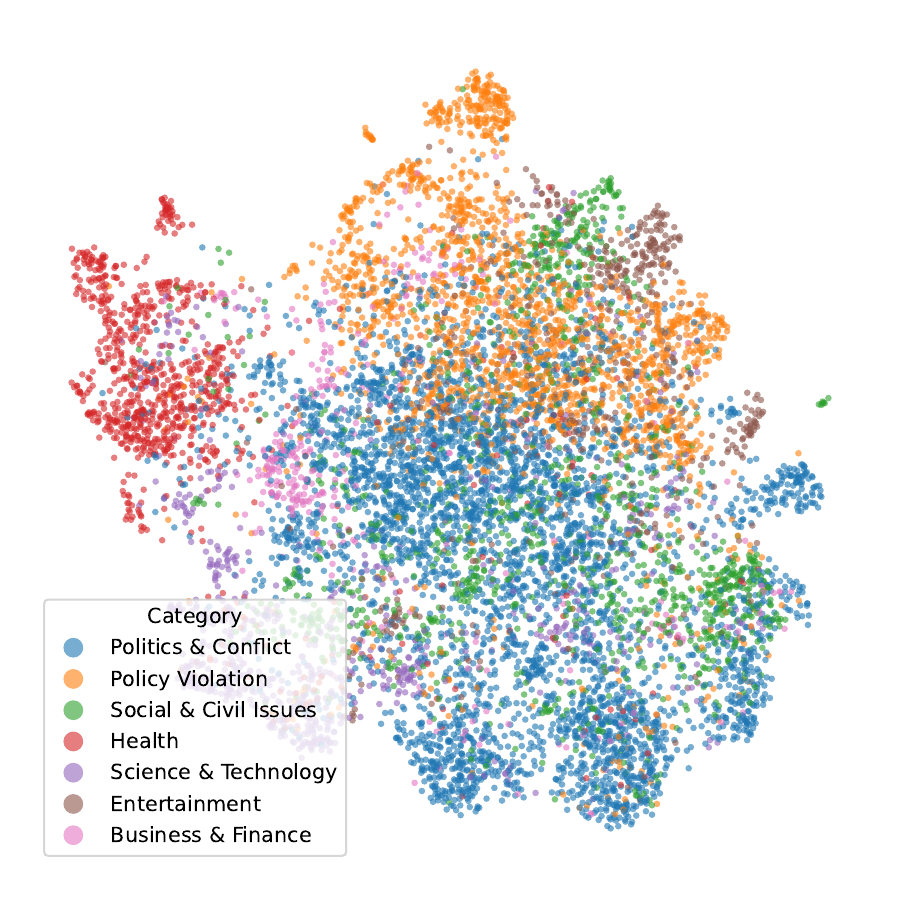}
    \caption{Visualization of topic modeling and axial coding results for 10,000 randomly selected \xs posts, colored by the seven broad categories. Posts belonging to the same category appear closer in the embedding space.}
    \Description{Visualization of topic modeling and axial coding results for 10,000 randomly selected X posts, colored by the seven broad categories. Posts belonging to the same category appear closer in the embedding space.}
    \label{fig:embeddings}
\end{figure}

\paragraph{Axial coding}
Two authors manually coded these 239 fine-grained topics based on the representative keywords and notes of each cluster. They began with the list of categories identified in prior literature~\cite{chuai2024did,pilarski2026supply}, while remaining open to proposing new categories during annotation. After one round of independent coding, all disagreements were resolved through discussion. Smaller categories were also merged into related categories. In the end, we identified seven broad categories as listed in~\Cref{tab:broad_category}. Politics \& Conflict is the largest category with 21,879 notes, followed by Policy Violation (13,373) and Social \& Civil Issues (6,705). The proportions of alerted notes also vary substantially across categories. Politics \& Conflict has the highest share of alerted notes (22.2\%), whereas Health has the lowest share (9.3\%). To examine the results, we projected the embeddings of 10,000 randomly selected \xs posts into a two-dimensional space as illustrated in~\Cref{fig:embeddings}. Visually, posts from the same category are positioned closer.

Policy Violation is a new category compared to prior work. It includes \xs posts related to altered media, clickbait, and scams. An AI-generated political post can be classified as both Politics \& Conflict and Policy Violation. However, while the political content is present in the original post, writers often omit such context in their notes. For example, a note claiming \textit{``this is a scam using AI-altered video''} explicitly identifies the type of misinformation but does not specify the underlying content. We acknowledge that it is a limitation of our approach, which relies on the concatenated note texts as input for topic modeling; however, this constraint arises from the unavailability of \xs post metadata.

\begin{figure*}[t]
    \centering
    \includegraphics[width=0.75\linewidth]{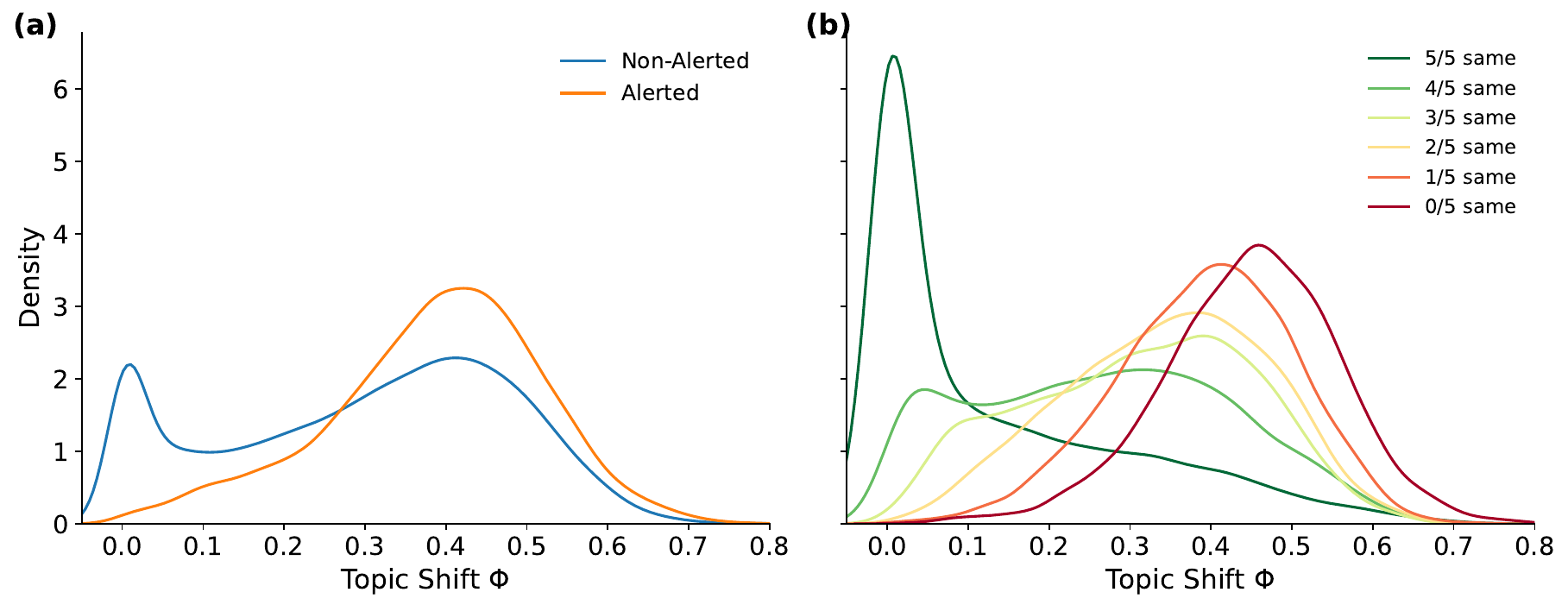}
    \caption{(a) Distributions of topic shift for alerted and non-alerted notes based on a balanced subsample in which every writer contributes the same number of notes. The bimodal pattern illustrates shifts within the same broad category (mode $\approx$ 0) and across different categories (mode $\approx$ 0.45). (b) Distributions of topic shift for all posts, stratified by the number of posts from the same broad category checked in the preceding five-note history (i.e., 5 $\rightarrow$ 0). The result shows that topic shift increases as the number of recently checked posts from the same category decreases.}
    \Description{(a) Distributions of topic shift for alerted and non-alerted notes based on a balanced subsample in which every writer contributes the same number of notes. (b) Distributions of topic shift for all posts, stratified by the number of posts from the same broad category checked in the preceding five-note history.}
    \label{fig:dist_topic_shift}
\end{figure*}

\paragraph{Quantifying topic shift}
In our setting, all notes associated with the same \xs post share the same embedding, fine-grained topic, and broad category, as these variables are intended to capture the topical interests reflected in the post being fact-checked. \swmn{\shortstack{SPC-4\\R3-4}}{For note $i$, we define ``topic shift'' $\Phi_i^k \in [0,1]$ as the cosine distance between its embedding ($\mathbf{e}_{i} \in \mathbb{R}^{768}$) and the average embedding of its writer's $k$ most recent notes ($\frac{1}{k}\sum^{k}_{j=1} \mathbf{e}_{i-j}$).} This metric quantifies how far a new note deviates from the writer's recent topical interests, with higher values indicating greater deviation. When constructing the $k$-note history, we include both alerted and non-alerted notes since both reflect the writer's interests. \sw{We also include notes with unavailable text as placeholders. Although the Note Status History dataset records their writer IDs and submission timestamps, the Notes dataset does not provide their text or associated \xs post IDs, preventing us from computing embeddings for these notes. Placeholder notes preserve the full temporal sequence of note-writing activity but are excluded when computing the average embedding of the $k$ most recent notes.} Cosine distance is a widely used measure of similarity between high-dimensional embeddings~\cite{anderson2020algorithmic,kim2025differential,hill2025pivot}. For non-alerted notes, topic shift captures the baseline variability of a writer's interests. For alerted notes, it reflects the extent to which platform prompts redirect writers toward content beyond their organic interests. \swmn{\shortstack{SPC-4\\R3-4}}{Comparing topic shift between these two groups allows us to assess the association between request alerts and the diversity of topics that writers fact-check. We report results for $k=5$ in the main paper. Robustness checks for $k=1,3,10$ are provided in~\Cref{app:k_topic_shift}.} We also provide a robustness check for early responders, defined as notes written in response to request alerts within 12 hours. The results are qualitatively similar~\Cref{app:t_alert_window}.

\subsection{Findings}
\label{ssec:diversity_findings}

\paragraph{Writers check more diverse posts under request alerts}
\swmn{R1-1}{Because the numbers of alerted and non-alerted notes vary substantially across writers, analyses based on all notes would implicitly assign greater weight to more prolific writers. To ensure that each writer contributes equally, we randomly sampled five alerted and five non-alerted notes per writer. This consideration also justifies our exclusion of writers with fewer than five notes in either group, as very sparse observations would provide limited writer-level information.} \Cref{fig:dist_topic_shift}(a) shows the resulting distributions of topic shift by alert status. Non-alerted notes exhibit a bimodal distribution: one mode near zero, corresponding to fact-checked posts that align closely with a writer's recent topical interests, and another around 0.45, corresponding to posts that differ substantially from those interests. This bimodal pattern suggests that writers naturally explore new topics even in the absence of request alerts. In contrast, alerted notes exhibit a largely unimodal distribution centered around 0.45. The near-zero mode disappears, indicating that notes written under request alerts are less likely to fall within a writer's recent areas of interest. \swmn{\shortstack{SPC-1a\\R3-1}}{Consistent with this interpretation, on average, alerted notes are associated with larger topic shifts than non-alerted notes.} \swmn{R1-1}{Results based on the full, unsampled dataset are reported in~\Cref{app:weighted_writer} as a robustness check. The results are qualitatively similar.}

\sw{To better understand the two modes, we plot the distributions of topic shift for all posts, stratified by the number of posts from the same broad category checked in the writer's five preceding notes in \Cref{fig:dist_topic_shift}(b). When all five prior posts belong to the same category, the distribution of $\Phi$ is centered near zero, indicating minimal topic shift. As the number of same-category posts decreases from five to zero, the distribution of $\Phi$ shifts away from zero. This confirms that larger $\Phi$ values capture cases in which the current post's topic differs from the writer's recent topical interests.}

\begin{figure*}[t]
    \centering
    \includegraphics[width=0.8\linewidth]{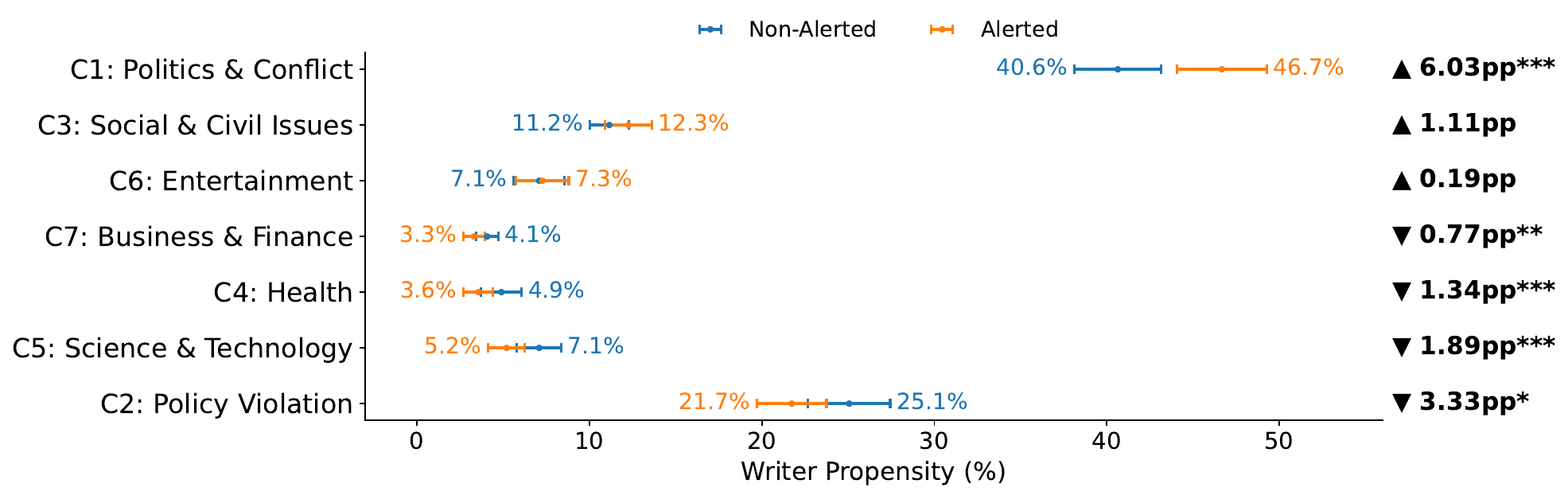}
    \caption{Writers' propensity to fact-check posts across broad categories, with and without request alerts. On average, a writer has a 46.7\% probability to check political content under alerts, increased from 40.6\% without alerts. Error bars indicate 95\% CIs. ``pp'' denotes percentage points. Statistical significance is assessed using the Wilcoxon signed-rank test, as each writer has a pair of observations in each group. $^{*}p < 0.05, ^{**}p < 0.01, ^{***}p < 0.001$.}
    \Description{Writers' propensity to fact-check posts across broad categories, with and without request alerts.}
    \label{fig:writer_propensity}
\end{figure*}

\paragraph{Writers check more political posts under request alerts}
For each writer, we computed the fraction of contributed notes in each broad category separately for the alerted and non-alerted groups, yielding a paired observation. We then averaged these fractions across different bootstrap samples of writers to obtain the mean and 95\% confidence intervals (CIs) of fact-checking propensity for each category. 

\Cref{fig:writer_propensity} presents the results. Politics \& Conflict category shows the largest increase in writer propensity, followed by Social \& Civil Issues and Entertainment. Compared to the non-alerted group, writers are more likely to fact-check political posts under request alerts (40.6\% $\rightarrow$ 46.7\%), corresponding to a 6.03 percentage point increase. The remaining four categories show declines in propensity. In particular, writers shift away from Policy Violation posts (25.1\% $\rightarrow$ 21.7\%), a decrease of 3.33 percentage points. Combined with the previous findings, these results suggest that although writers tend to fact-check more diverse content under alerts at the individual level, the overall Community Notes ecosystem does not become more diverse. Instead, contributions collectively shift toward political content, increasing concentration in this already dominant category. The share of Politics \& Conflict notes rises from 38.3\% among non-alerted notes to 46.6\% among alerted notes, as calculated using data from~\Cref{tab:broad_category}. 

\begin{figure*}[t]
    \centering
    \includegraphics[width=1\linewidth]{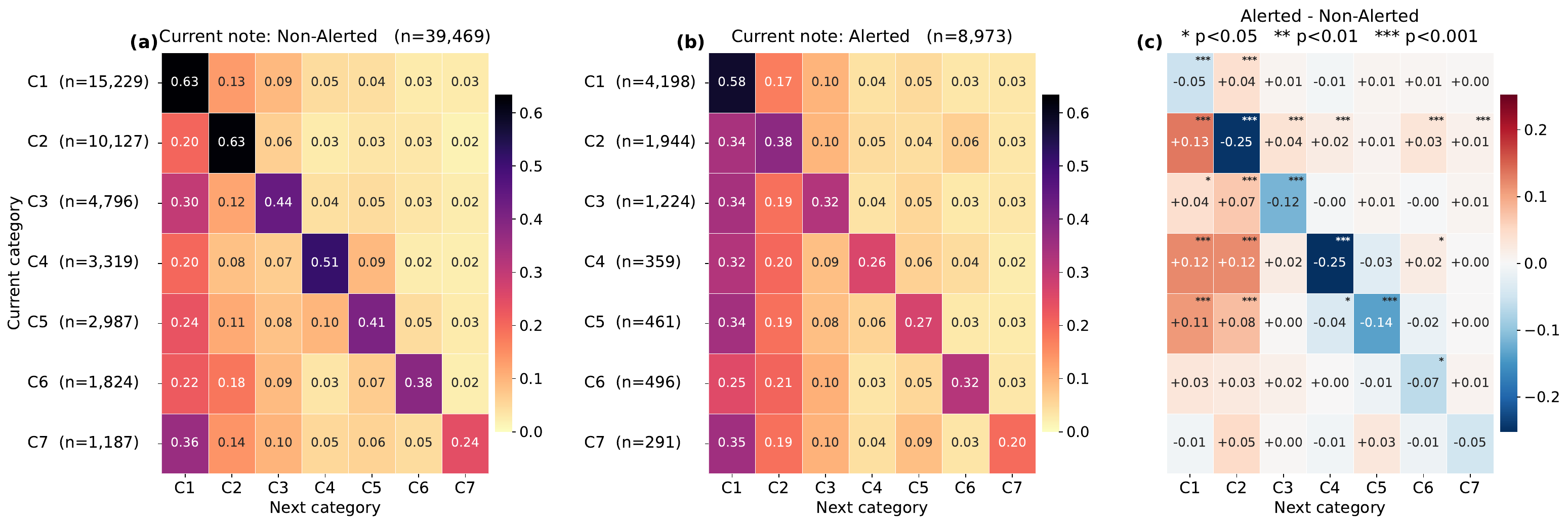}
    \caption{Category-to-category transition matrices for non-alerted notes (a) and alerted notes (b), along with their difference (c). Rows represent the category of the current post (sample sizes shown in parentheses), and columns represent the category of the subsequent post. Cell values indicate the probability of transitioning between categories, such that each row sums to one.} 
    \Description{Category-to-category transition matrices for non-alerted notes (a) and alerted notes (b), along with their difference (c).}
    \label{fig:topic_shift_category}
\end{figure*}

\swmn{R3-6}{\Cref{fig:topic_shift_category}(a) shows the category-to-category transition matrix for non-alerted notes. Each row corresponds to the category of the current post, with sample sizes reported in parentheses, and each column corresponds to the category of the subsequent post. Cell values indicate transition probabilities, such that each row sums to one. We find that writers are generally most likely to continue fact-checking posts within the same category, as evidenced by the darker cells along the diagonal, except for C7: Business \& Finance. \Cref{fig:topic_shift_category}(b) shows the corresponding matrix for alerted notes. Relative to the non-alerted group, transitions toward C1: Politics \& Conflict and C2: Policy Violation become more frequent across a wide range of source categories, as reflected by the darker cells in the first two columns.}

\sw{This pattern is more apparent in \Cref{fig:topic_shift_category}(c), which shows the difference between the alerted and non-alerted transition matrices. Compared with the non-alerted baseline, writers are less likely to remain within the same category and more likely to transition to C1 and C2 when writing alerted notes. These results are not contrary to \Cref{fig:writer_propensity}, which shows the largest overall decline in the propensity of Policy Violation category. Although alerted notes are more likely to transition into Policy Violation from other categories, Policy Violation also accounts for a large baseline share, and the reduction in within-category transitions for Policy Violation itself is substantial ($-0.25^{***}$), resulting in a net decline in overall propensity. Statistical significance is assessed using two-sided two-proportion $z$-tests that compare each transition probability under alerts with its non-alerted counterpart, with Benjamini--Hochberg correction applied across all 49 cells.}


\swmn{\shortstack{SPC-1a\\R3-1}}{Overall, these findings suggest that request alerts are associated with contributors writing notes for a broader set of posts with respect to their prior interests, highlighting their relevance as interface cues in community-requested fact-checking. Nevertheless, this diversification is not uniform across categories. Instead, alerted notes are disproportionately associated with political content, consistent with Community Notes' original focus on political fact-checking.}


\section{Association with Note Visibility}
\label{sec:visibility}

Community Notes become visible to the public only when they attain the CRH status. Notes that fail to reach this status remain hidden and are visible only to enrolled participants to rate. Therefore, CRH status serves as the primary indicator of note visibility. The proportion of CRH notes reflects the effectiveness of Community Notes in combating misinformation at scale. In this section, \swmn{\shortstack{SPC-1a\\R3-1}}{we examine how the request alert status is associated with the CRH probability of notes across different categories.}

As described earlier, there are two alternative CRH metrics: currently CRH and ever CRH. We choose to model the ever CRH metric for two reasons. First, a note that is currently CRH at the time of data collection may still lose this status in the future, reflecting the instability of note publication status documented in~\cite{chuai2026consensus}. Second, ever CRH captures whether a note is visible to the public at any point in its lifetime, better aligning with the definition of visibility. We also experiment with the currently CRH metric and report the results in~\Cref{app:current_crh}. The results are qualitatively similar.

\subsection{Methods}
\label{ssec:visibility_methods}

The rightmost columns in~\Cref{tab:broad_category} show the ever CRH rates stratified by alert status. We observe that the ever CRH rates of alerted notes are higher than these of non-alerted notes across all categories. Some categories (e.g., Science \& Technology and Policy Violation) appear to reach consensus more easily than others (e.g., Politics \& Conflict). However, we caution that, while category-level CRH rates provide an intuitive comparison between the two groups, they do not constitute a rigorous measure. In particular, some categories may attract higher quality writers, thereby inflating the observed CRH rates. For example, writers who fact-check Science \& Technology posts may possess domain expertise, enabling them to better identify misleading content and consistently submit higher quality notes. The observed increase in the ever CRH rate, from 20.64\% for non-alerted notes to 42.39\% for alerted notes, may be attributed to that such domain experts contribute more frequently in response to requests. To ensure fair estimates, an appropriate model must account for heterogeneity across both writers and fine-grained topics. We therefore employ a mixed-effects logistic regression model, as specified below.
\begin{equation*}
\mathrm{logit}\!\big(Pr(Y_i)\big)
= \beta_0 + \beta_1 A_i + \boldsymbol{\beta}^\top \mathbf{C}_i
+ \boldsymbol{\gamma}^\top (A_i \mathbf{C}_i)
+ w_{i} + t_{i}
\end{equation*}

The model specifies the log-odds of the outcome $Y_i = 1$ for note $i$. Here, $Y_i$ is a binary indicator of whether note $i$ ever attains the CRH status. $Pr(Y_i)$ is the predicted CRH probability. $A_i$ denotes whether note $i$ is written under an alert. $C_i$ is the broad category of note $i$. Both $A_i$ and $C_i$ are modeled as fixed effects. $A_i \mathbf{C}_i$ captures the interaction effects between alert status and broad category. We include random intercepts for both writers and fine-grained topics. $w_{i}$ is the writer-level random effects, capturing unobserved heterogeneity across writers (e.g., some writers consistently produce higher quality notes). $t_{i}$ is the topic-level random effects, capturing heterogeneity across fine-grained topics (e.g., some topics are easier to reach consensus). $\beta_0$ is the baseline log-odds of a note being ever classified as CRH. 

Mixed-effects models assume that random effects are drawn from a common population distribution and are independent of the residual errors. In our setting, writers and topics are naturally treated as random effects, as they represent samples from a broader population of potential writers and topics. In contrast, fixed effects correspond to systematic predictors of interest whose effects are estimated directly. To ensure a stable and interpretable representation, we map the 239 fine-grained topics into a fixed set of seven broad categories via axial coding. By including random intercepts for writers and fine-grained topics, the mixed-effects model accounts for unobserved heterogeneity across these dimensions. This specification enables us to estimate the difference between the expected CRH probabilities for a random writer addressing a random fine-grained topic from each broad category, with and without request alerts. 

\swmn{R1-2}{One possible extension is to include post-level random intercepts to account for unobserved heterogeneity across target posts. This could help capture post-specific factors, such as post visibility, popularity, or author prominence, that may be associated with both alert status and CRH probability. However, this specification is not well supported by our data because each post has, on average, only 1.07 notes. With so few repeated observations per post, a post-level random effect would be weakly identified and could lead to unstable estimates. We therefore only include writer- and topic-level random intercepts in the mixed-effects model.}

\subsection{Findings}
\label{ssec:visibility_findings}

\begin{figure*}[t]
    \centering
    \includegraphics[width=0.8\linewidth]{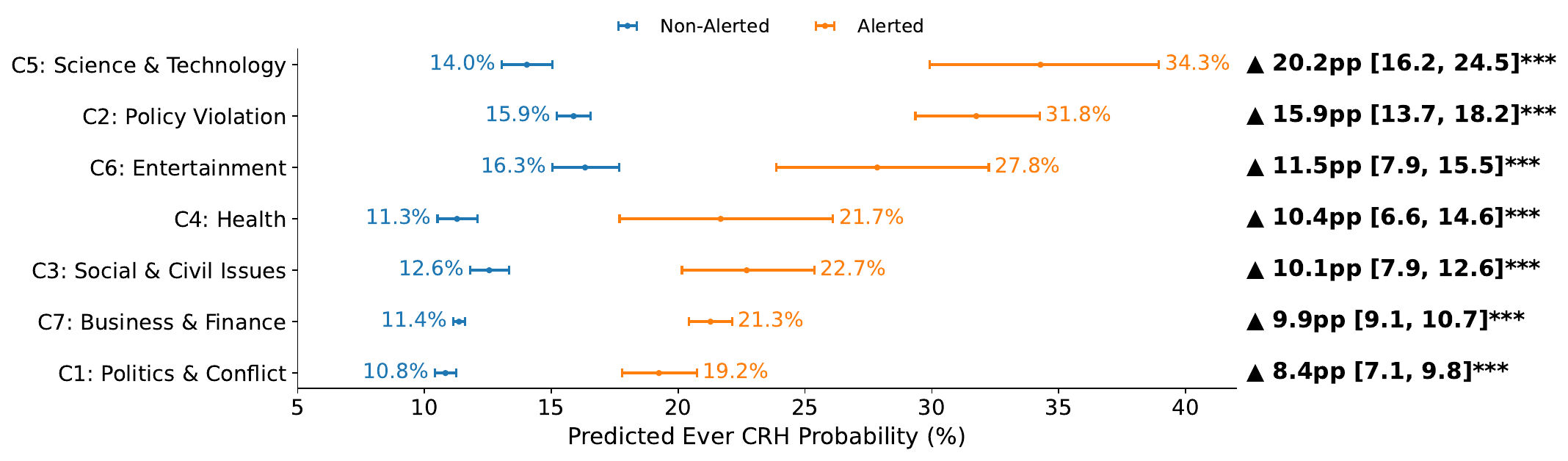}
    \caption{Association between alert status and note visibility across categories. The mean and 95\% CIs of predicted ever CRH probabilities are estimated from a mixed-effects logistic regression model with random intercepts for writers and fine-grained topics. Numbers in brackets denote the 95\% CIs for the increase of predicted ever CRH probability, estimated via bootstrapping. Statistical significance is assessed using the Wilcoxon signed-rank test. $^{***} p < 0.001$.}
    \Description{Association between alert status and note visibility across categories.}
    \label{fig:mix_effect}
\end{figure*}

\paragraph{Request-alerted notes are more likely to be visible}
\Cref{fig:mix_effect} shows that the presence of request alerts is associated with significantly higher ever CRH probabilities across all categories. The largest increase is observed in the Science \& Technology category. For a random writer checking a random Science \& Technology post, if the post has no alert, the model estimates that the resulting note has 14.0\% (95\% CIs: [13.0\%, 15.0\%]) probability to reach CRH status at some point. The ever CRH probability increases to 34.3\% (95\% CIs: [29.8\%, 38.9\%]) if the post has a request alert. This corresponds to a 20.2 (95\% CIs: [16.2\%, 24.5\%]) percentage point increase. Even the smallest increase, observed in the Politics \& Conflict category (10.8\%$\rightarrow$19.2\%), amounts to 8.4 percentage points.

\begin{figure}[t]
    \centering
    \includegraphics[width=0.55\linewidth]{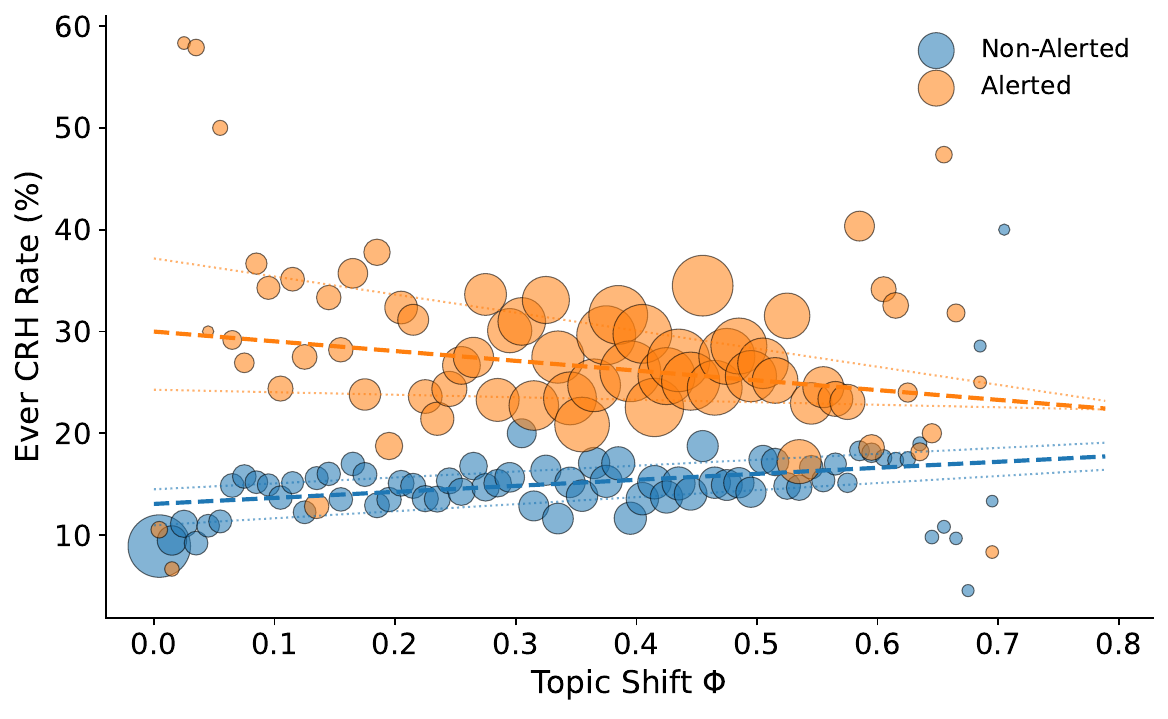}
    \caption{The relationship between topic shift and CRH rate. Circle: notes written with such a topic shift value, sized by the number of notes. Lines are fitted by quantile regression to mitigate imbalanced data distribution along the x-axis. Dashed lines indicate fitting the median values. Dotted lines are 95\% CIs.}
    \Description{The relationship between topic shift and CRH rate.}
    \label{fig:crh_vs_topic_shift}
\end{figure}

\paragraph{Visibility gain diminishes with topic shift}
\Cref{fig:crh_vs_topic_shift} shows the relationship between topic shift and the ever CRH rate, separately for alerted and non-alerted notes. The x-axis is binned at 0.01 intervals of topic shift. \swmn{R3-7}{The data are unevenly distributed along the x-axis, with higher density, for example, around $\Phi$ = 0 and between 0.3 and 0.5. This pattern matches the topic shift distribution shown in~\Cref{fig:dist_topic_shift}(a).} As a result, a simple linear regression would be heavily influenced by these high-density regions. We therefore employ quantile regression for its robustness to such imbalanced distributions. We find that, for non-alerted notes, ever CRH rates remain relatively stable, with a slight increase as topic shift increases (slope = +2.55). In contrast, under request alerts, the gains diminish as writers move further away from their recent topical interests (slope = -8.96). However, visibility remains consistently higher for alerted notes compared to their non-alerted counterparts.


\section{Discussion}
\label{sec:discussion}

\subsection{Individual Exploration, Collective Inequality}
Request-alerted notes are associated with broader topical engagement at the individual writer level, but this pattern coexists with a counterintuitive collective pattern. Although contributors write notes on more diverse topics relative to their own prior activity, alerted notes are collectively more concentrated in a narrower set of topics, most notably Politics \& Conflict. At the same time, comparatively ``softer'' categories, such as Science \& Technology, are less represented among alerted notes. In other words, individual-level topical diversity does not necessarily correspond to system-level diversity; instead, alerted notes are disproportionately concentrated in already dominant domains. \swmn{R1-3}{This pattern differs from findings in algorithmic recommendation systems, where reduced diversity has been observed when users consume recommended content surfaced through system interface cues~\cite{anderson2020algorithmic}.}

This divergence can be understood through the lens of incentive alignment and coordination in collective systems. In the absence of request alerts, contributors may rationally prioritize topics that are easier to evaluate and more likely to yield visible outcomes, such as achieving consensus. Highly polarized political content, by contrast, is more difficult to adjudicate and often requires greater coordination, making it less attractive despite its societal importance. By signaling unmet demand~\cite{pilarski2026supply}, request alerts alter this calculus by directing attention toward tasks that are individually costly but collectively valuable. In this sense, request alerts function as an endogenous coordination mechanism, aligning contributor effort with socially prioritized yet otherwise under-supplied areas of fact-checking.

However, since demand is itself shaped by user attention, this coordination mechanism can amplify existing imbalances. Topics that are already salient or contentious generate more requests and thus attract disproportionate contributor effort, while less visible but equally important issues, such as consumer protection or health misinformation, remain underrepresented. As a result, a purely demand-driven system may exacerbate aggregate inequality even as it promotes individual exploration. Addressing this tension requires complementing request-based signals with supply-side interventions, such as targeted task routing or algorithmic promotion of under-served topics, to ensure a more balanced allocation of attention across the information ecosystem.

\subsection{Why Do Request Alerts Work?}

\swmn{\shortstack{SPC-1\\R3-1}}{Our results show that notes written for posts with request alerts are associated with greater note visibility, even though these posts appear to involve more challenging and contested content. This pattern is not consistent with a simple explanation in which alerted notes are more visible because they are attached to easier tasks. Instead, it suggests that alerted posts may represent a distinct set of contribution opportunities: they are publicly signaled as needing attention while also involving topics that may be more difficult to evaluate.} This pattern can be understood through the lens of interface cue design and collective action. Prior research on the MAIN model~\cite{sundar2008main} and heuristic processing suggests that users rely on salient cues to allocate attention under limited cognitive resources~\cite{kim2011using}. In this context, request alerts function as hybrid cues that combine visual salience with social signals: they make community-requested posts more visible to eligible contributors while also indicating that other users perceive the post as needing additional context.

\sw{At the same time, alerted posts also appear to be associated with more ambiguous, contested, and difficult-to-evaluate content. Research on misinformation correction shows that such claims are more persistent and resistant to debunking~\cite{chan2017debunking}, implying that they require greater effort and coordination to resolve. Prior work also finds that prompting individuals to search for or review evidence improves misinformation judgments and reduces partisan bias~\cite{resnick2023searching}. Request alerts may activate a similar mechanism. By signaling community concern, they encourage contributors to engage more deeply with evidence, even in difficult cases. Although our observational analysis cannot determine whether request alerts change contributors' behavior or effort, our findings are consistent with the possibility that community requests help surface collectively important cases that contributors may not otherwise prioritize.}


\subsection{The Pivot Penalty}

We find that the visibility gains associated with request alerts are not uniform; rather, they diminish as the topic diverges from a contributor's prior areas of engagement. Although alerts encourage contributors to explore new topics, notes produced further from a contributor's historical focus are less likely to achieve CRH status. This pattern suggests that directing attention alone is insufficient, as the effectiveness of request alerts also depends on how well the prompted task aligns with a contributor's existing knowledge.

This pattern mirrors the ``pivot penalty'' identified in recent research on scientific production, where individuals who move into new domains experience declines in impact~\cite{hill2025pivot}. In our setting, contributors who pivot to unfamiliar topics may face similar constraints, including limited domain knowledge, greater uncertainty, and higher coordination costs in reaching consensus. While request alerts successfully induce exploration, they also introduce performance costs when contributors move too far from their areas of expertise. This highlights a trade-off between exploration and effectiveness in crowdsourced fact-checking systems: expanding the scope of participation may come at the expense of contribution quality and consensus formation.

\subsection{Design Implications}

From a design perspective, these findings suggest that the effectiveness of request alerts lies not merely in increasing engagement, but in structuring how attention and effort are distributed across tasks. Alerts act as coordination mechanisms that can redirect contributors toward socially important but under-addressed content. However, because demand signals are shaped by user attention, they may also reinforce existing imbalances, concentrating effort on already salient topics. Designing effective systems therefore requires going beyond demand-driven signaling to more deliberately guide participation. For example, platforms could prioritize tasks that are unlikely to reach consensus without additional input or that are systematically underrepresented, ensuring that collective effort is allocated more evenly across the system.

Our findings on the pivot penalty also highlight the importance of aligning task assignment with contributor expertise. While alerts encourage exploration, their benefits are greatest when contributors engage with topics that are adjacent to their existing expertise. This points to the value of more targeted interventions, such as matching alerts to contributors based on prior activity or recommending tasks within nearby topical neighborhoods. Incorporating signals of both task difficulty and contributor specialization into system design may help mitigate performance costs while preserving the benefits of diversified engagement. Broadly, platforms that jointly optimize attention allocation and expertise matching may better leverage crowd-based fact-checking to improve both the quality and coverage of misinformation correction.

\subsection{Limitations and future work}
This study has several limitations related to data availability and measurement. To ensure transparency and reproducibility, we restrict our analysis to publicly available Community Notes data and do not incorporate the underlying \xs post content or platform-level metadata. As a result, we infer topical information from concatenated note texts rather than from the original posts themselves. While this approach captures how contributors describe and contextualize content, it may not fully reflect the original semantics or intent of the \xs posts being fact-checked. More importantly, it limits our ability to directly characterize the difficulty of reaching consensus across posts, which is an important factor in understanding the effectiveness of request alerts.

In particular, our findings cannot fully disentangle two potential mechanisms underlying the observed visibility gains. On one hand, request alerts may redirect contributors toward posts that are more likely to contain misinformation and thus more likely to reach consensus once sufficiently scrutinized. On the other hand, alerts may improve outcomes by inducing higher-quality contributions, for example by increasing effort or facilitating coordination. Distinguishing between these mechanisms requires richer data on post content, claim verifiability, and consensus dynamics. Future work that integrates Community Notes data with platform-level signals could explicitly model the difficulty of fact-checking tasks and examine how alerts interact with contributor expertise and content characteristics. Such work would provide a more precise understanding of whether request alerts operate primarily through task selection, contribution quality, or their interaction.

\swmn{SPC-2}{
The criteria for displaying request alerts have evolved over time. In this study, we infer historical alert status based on the Community Notes documentation available at the time of data collection. Because the criteria changed after our data collection date of June 26, 2025, and the threshold for displaying request alerts became easier to satisfy, more \xs posts may receive request alerts under the current system than under the criteria reflected in our dataset. As a result, recent Community Notes activity may include a larger share of contributions from top writers that are aligned with community requests. This change may affect the generalizability of our findings: the associations we observe characterize request alerts under the criteria in place during our study period, and future work should examine whether similar patterns hold under the newer, broader alerting criteria.
}

\sw{Finally, because our analyses are based on observational data, the observed associations may reflect selection or other unobserved differences between posts with and without request alerts, and therefore should not be interpreted as causal effects of request alerts. Although Community Notes' scoring system reduces one form of strategic manipulation---labels on deleted notes continue to affect contributor scores, preventing writers from trivially avoiding penalties by deleting CRNH notes---other forms of selection may remain. In particular, writers may strategically choose posts or categories that are easier to judge, such as Policy \& Violation, to improve their writing impact. Such self-selection could influence both the types of posts writers engage with and the likelihood that their notes are modeled as helpful. We therefore interpret the observed differences in note visibility and topical engagement as associations rather than evidence that request alerts directly change contributor behavior or note outcomes.}

\paragraph{Ethical considerations}
This study relies exclusively on publicly available Community Notes data and does not involve interaction with human subjects. Generative AI tools were used to assist with writing and programming. The authors retain full responsibility for all analyses, interpretations, and the content of the manuscript.


\section{Conclusion}

In this work, we examine request alerts as an interface cue in Community Notes and identify a nuanced set of associations across individual topical engagement, collective content coverage, and note visibility. First, at the individual level, notes written for posts with request alerts are associated with greater topical diversity and more political content, relative to writers' prior activity. Second, at the aggregate level, these individual-level differences coexist with stronger concentration in the already dominant Politics \& Conflict category, suggesting greater content inequality within the system. Third, alerted notes are substantially more likely to be classified as helpful and to become publicly visible, indicating a positive association between alert status and note visibility. Finally, this visibility gain is not uniform: it is smaller when contributors write about topics farther from their prior interests, revealing a pivot penalty associated with topical divergence. Altogether, these findings suggest that request alerts are associated with distinct patterns of contributor activity and note visibility in crowdsourced fact-checking systems, while highlighting important trade-offs between exploration, expertise, and equitable content coverage.

\bibliographystyle{ACM-Reference-Format}
\bibliography{references}

\appendix

\section{Robustness Checks}
\label{app:robust_checks}

\subsection{Varying Embedded Objects: Post Embedding vs. Concatenated-Note Embeddings}
\label{app:post_embedding}

\begin{figure}[h]
    \centering
    \includegraphics[width=0.55\linewidth]{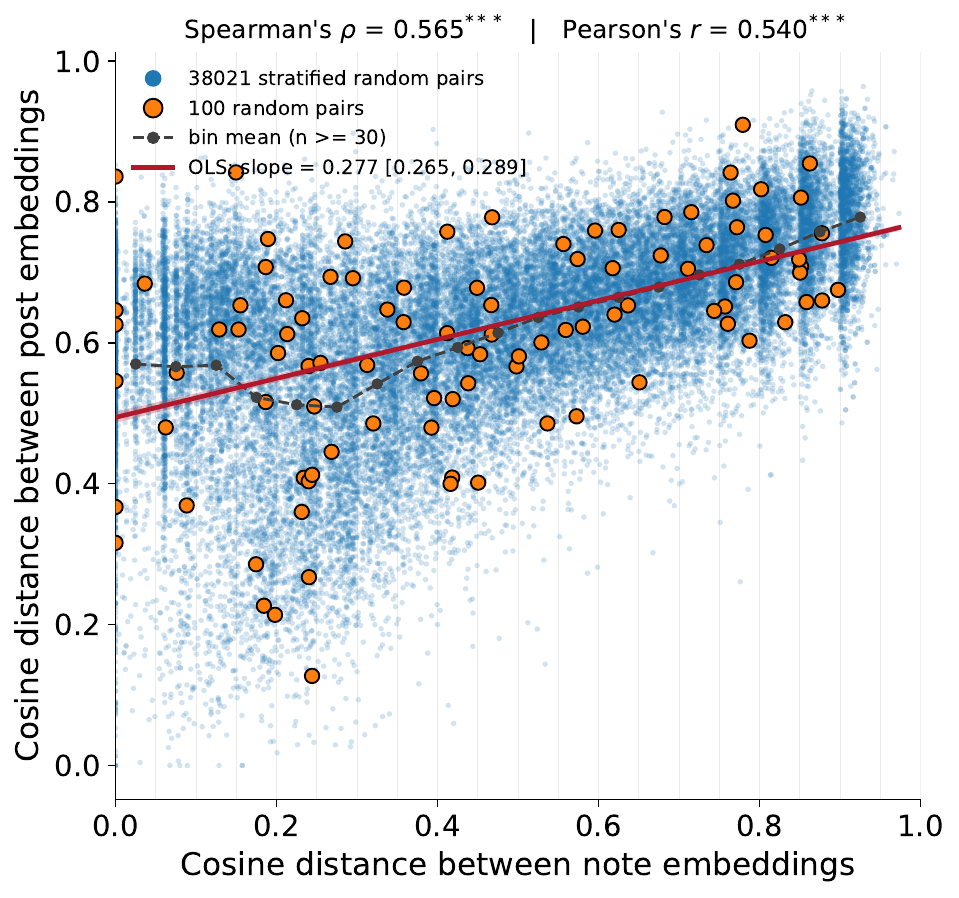}
    \caption{Cosine distance between concatenated-note embeddings versus between post embeddings, for randomly sampled pairs of posts stratified into 20 equal-width bins of note distance. Dashed line: bin means. Red line: OLS fit with a 95\% band from standard errors clustered on both posts.}
    \Description{Cosine distance between concatenated-note embeddings versus between post embeddings.}
    \label{fig:app_post_vs_note}
\end{figure}

\Cref{fig:app_post_vs_note} shows the positive correlation between concatenated-note embeddings and post embeddings. We split x-axis into 20 bins. We iteratively sample a pair of posts, compute the cosine distance between their concatenated-note embeddings, and then place the pair into corresponding bin. Each bin contains up to 200 pairs. For each sampled pair, we scrape their \xs post metadata and compute the cosine distance between their post embeddings. The results show that cosine distances calculated from concatenated-note embeddings are significantly positively correlated with those calculated directly from post embeddings on a sample of \xs posts (Spearman's $\rho = 0.567, p<0.001$, Pearson's $r = 0.534, p<0.001$), supporting the use of concatenated-note embeddings as a proxy for post content.

\subsection{Varying Length of Most Recent Note History: 1, 3, 5, 10}
\label{app:k_topic_shift}

\begin{figure*}[h]
    \centering
    \includegraphics[width=1\linewidth]{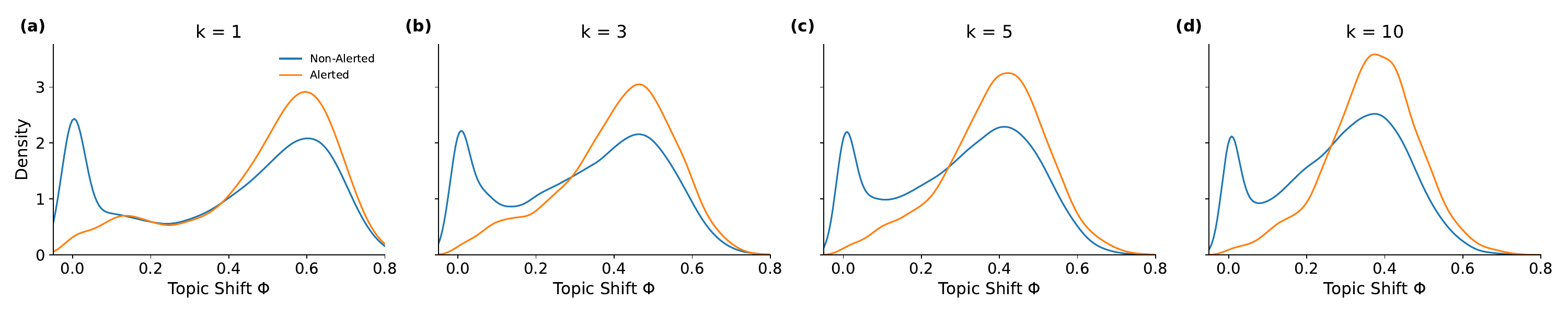}
    \caption{Varying the length $k=1,3,5,10$ used to compute a writer's prior topical interest.}
    \Description{Varying the length $k=1,3,5,10$ used to compute a writer's prior topical interest.}
    \label{fig:app_varying_k}
\end{figure*}

We vary the length $k$ used to compute a writer's prior topical interest and replicate the analyses in~\Cref{ssec:diversity_findings}. The results, as shown in~\Cref{fig:app_varying_k}, are all qualitatively similar: writers check more diverse posts under request alerts.

\clearpage

\subsection{Comparing with Early Responders: 12, 24}
\label{app:t_alert_window}

\begin{figure*}[h]
    \centering
    \includegraphics[width=0.55\linewidth]{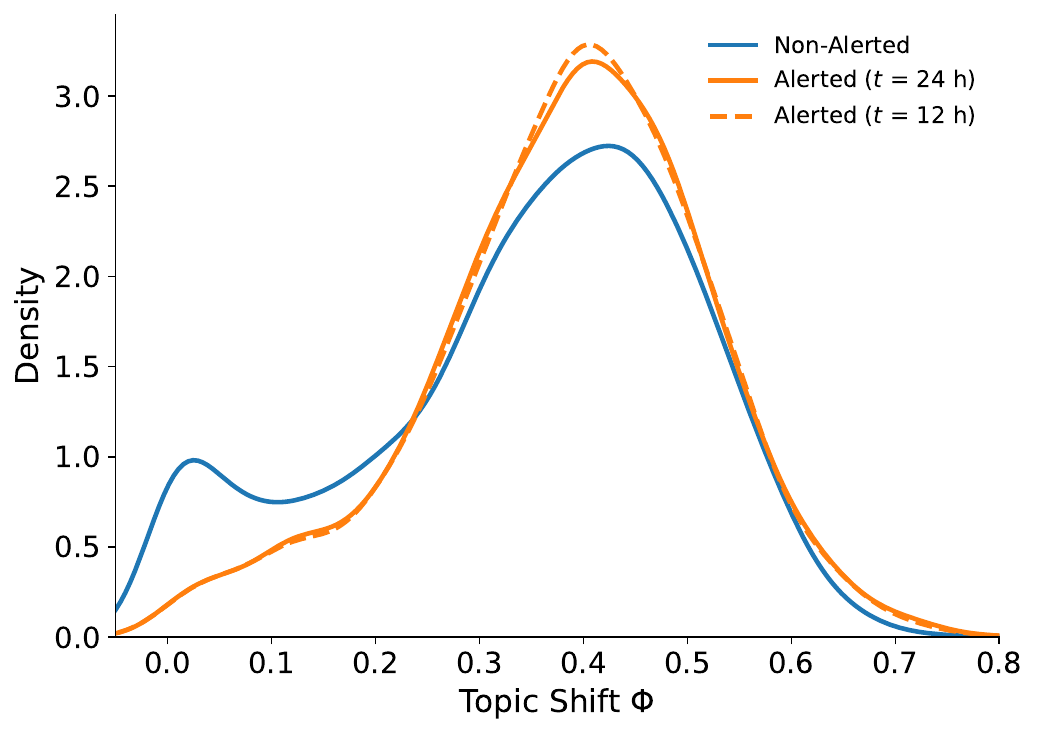}
    \caption{Varying the time window $t=12,24$ used to separate early responders and all alert responders.}
    \Description{Varying the time window $t=12,24$ used to separate early responders and all alert responders.}
    \label{fig:app_varying_t_window}
\end{figure*}

We extract all notes written within 12 hours of the 5th request. This separates a group of early responders. The results, as shown in~\Cref{fig:app_varying_t_window}, are all qualitatively similar.

\subsection{Varying Writer Weights: Equal Weights for All Writers vs. More Weights for Prolific Writers}
\label{app:weighted_writer}

\begin{figure}[h]
    \centering
    \includegraphics[width=0.55\linewidth]{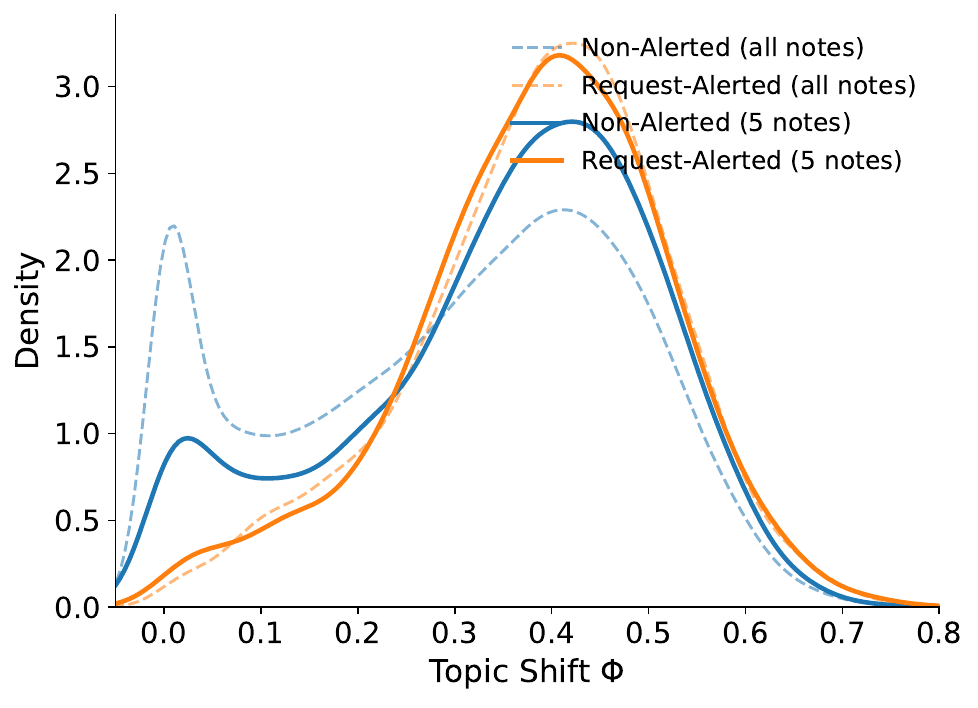}
    \caption{Assigning equal weights to all writers vs. assigning more weights to prolific writers.}
    \Description{Assigning equal weights to all writers vs. assigning more weights to prolific writers.}
    \label{fig:varying_weights}
\end{figure}

In contrast to randomly sampling five alerted and non-alerted notes to calculate the writer-level topic shifts, we use all notes and show the results in~\Cref{fig:varying_weights}. The results are qualitatively similar: writers check more diverse posts under request alerts.

\section{Associations with Currently CRH Metric}
\label{app:current_crh}

We use an alternative visibility metric, Currently CRH, as the output variable in the mix-effects model as described in~\Cref{ssec:visibility_methods}. The results are qualitatively similar: notes written under alerts are significantly more visible, though the degrees differ with estimating used the Ever CRH metric.

\begin{figure}[h]
    \centering
    \includegraphics[width=0.8\linewidth]{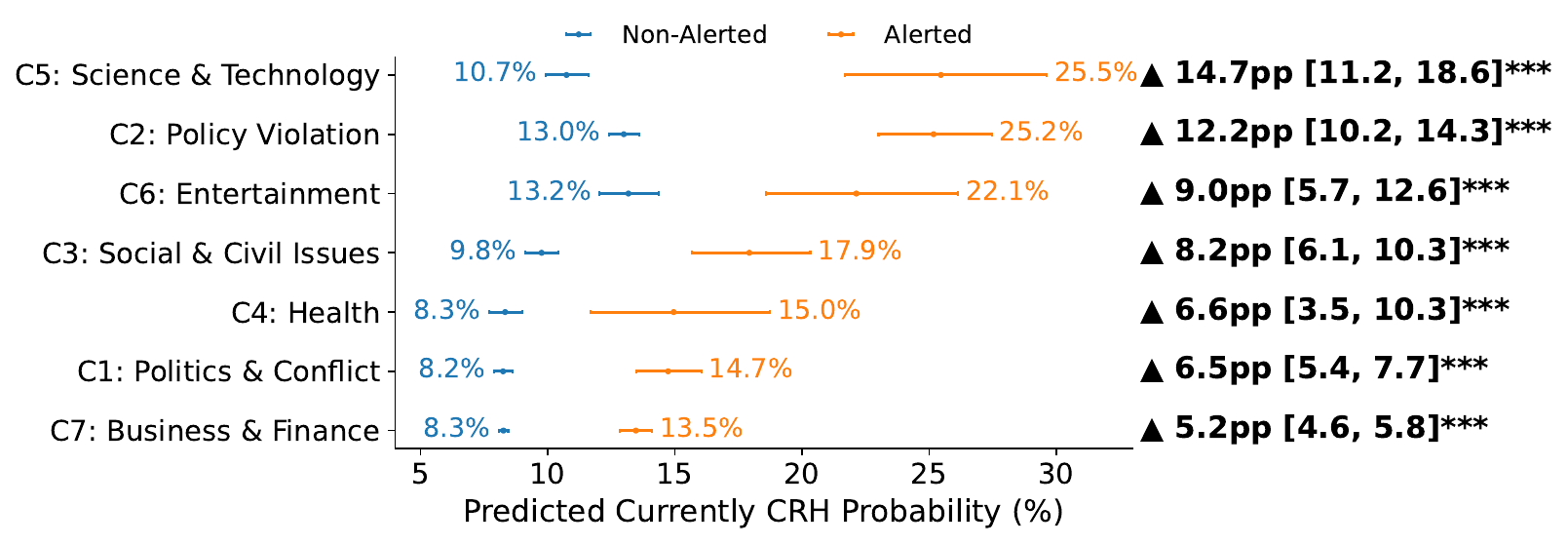}
    \caption{The presence of request alerts is significantly associated with higher currently CRH probabilities across all categories, consistent with our findings in~\Cref{ssec:visibility_findings}.}
    \Description{The presence of request alerts is significantly associated with higher currently CRH probabilities across all categories.}
    \label{fig:app_mix_effect}
\end{figure}

\end{document}